\documentclass[12pt]{article}

\usepackage{amssymb}

\usepackage{graphicx}

\advance \topmargin by -\headheight

\advance \topmargin by -\headsep

\evensidemargin \oddsidemargin

\newcommand{\beq}{\begin{equation}}

\newcommand{\eeq}{\end{equation}}

\newcommand{\bear}{\begin{eqnarray}}

\newcommand{\eear}{\end{eqnarray}}

\begin{document}

\begin{titlepage}

\begin{center} \Large \bf Nuclear like effects in proton-proton collisions \\at high energy

\end{center}

\vskip 0.3truein

\begin{center}

L. Cunqueiro${}^{\,1}$, J. Dias de
Deus${}^{\,2}$ and C. Pajares${}^{\,1}$

\vspace{0.3in}

${}^{\,1}$Departamento de F\'\i sica de Part\'\i culas, Universidade de
Santiago de Compostela \\and Instituto Galego de F\'\i sica de Altas Enerx\'\i
as (IGFAE), \\ E-15782 Santiago de Compostela, Spain.

${}^{\,2}$CENTRA, Instituto Superior T\'ecnico, 1049-001 Lisboa, Portugal.

\end{center}

\vspace{1cm}

\begin{center}

\bf ABSTRACT

\end{center}

We show that several effects considered nuclear effects are not nuclear in
the sense that they do not only occur in nucleus-nucleus and hadron-nucleus
collisions but, as well, they are present in hadron-hadron (proton-proton)
collisions. The matter creation mechanism in $hh$, $hA$ and $AA$ collision is always the
same. The $p_{T}$ suppression of particles produced in large multiplicity
events compared to low multiplicity events, the elliptic flow and the Cronin
effect are predicted to occur in $pp$ collisions at LHC energies as a
consequence of the obtained high density partonic medium.

\vskip5.6truecm

\smallskip

\end{titlepage}

\setcounter{footnote}{0}

The ``nuclear effects'' considered in high energy strong
interacting physics are essentially not nuclear but resulting from the partonic
medium dictated by the underlying theory, QCD. The physics in central $pp$ collisions is
not physics in vacuum but physics in a medium whose properties are
universal. At very high energy the wave function of the initial protons
contains a high number of partons in such a way that the partons of each
proton probe a high density medium. As we shall see, the ``nuclear modification factor''
$R_{AB}$ \cite{RHIC} or the distribution ratio central/peripheral,
$R_{CP}$\cite{RHIC}, and the ``elliptic flow'' parameter $v_{2}$
\cite{Adler:2003kt}\cite{starv2}, for instance, can be applied to $pp$ collisions.

Note that to understand $hA$ and $AA$ collisions we need theoretical high
energy nuclear physics, namely Glauber calculus \cite{Glauber:1970jm}. Without it we
could not estimate cross-sections, rescattering effects, the number of binary
collisions or the number of
participating nucleons. Of course, in $pp$ collisions $N_{coll}=1$,
$N_{part}=1$.

We formulate our arguments in the frame of the string percolation model
\cite{Armesto:1996kt}, but we think that the same formulation is valid in the Color Glass
Condensate model \cite{McLerran:1993ni}. Both of them are able to reproduce most of the RHIC
data, obtaining a similar transverse momentum scale \cite{SchaffnerBielich:2001qj}\cite{DiasdeDeus:2003ei}; the saturation momentum
scale $Q_{s}$ and the percolation threshold are related to each other
\cite{DiasdeDeus:2006xk}. The discussed effects are a consequence of the high
density partonic medium formed in  the collision, independently of the detailed framework used for the description.

The similarities of $hh$ and $AA$ collisions were previously explored two
decades ago \cite{Von Gersdorff:1986yf}\cite{McLerran:1986nc} studying a hydrodynamical description of $AA$ and
$\overline{p} p$ fluctuations and the dependence of the mean transverse
momentum on the multiplicity. 

In the string percolation model, multiparticle production is described in
terms of color strings stretched between the partons of the projectile and
the target. These strings decay into new ones by $q \overline{q}$ or
$qq-\overline{q}\overline{q}$ pair production and subsequently hadronize to
produce the observed hadrons. Due to confinement, the color of these strings
is confined to a small area in transverse space $S_{1}=\pi r_{0}^{2}$, with
$r_{0} \simeq 0.2-0.3$ fm. With increasing energy and/or atomic number of the
colliding particles, the number of strings $N_{s}$ grows and they start to
overlap forming clusters, very much like disks in two-dimensional percolation
theory. At a certain critical density, a macroscopical cluster appears, which
marks the percolation phase transition \cite{pajares} \cite{Pajares:2005kk}. This density
corresponds to the value of $\eta=N_{s}\frac{S_{1}}{S_{A}}$, $\eta_{c}=1.2$, where 
$S_{A}$
stands for the overlapping area of the colliding objects. A cluster of $n$
strings behaves as a single string with energy-momentum corresponding to the
sum of the individual ones and with a higher color field corresponding to the
vectorial sum in color space of the color fields of the individual strings. In
this way, the mean multiplicity $ \langle \mu_{n} \rangle $ and the mean transverse momentum squared
$ \langle {p_T}^{2}_{n} \rangle$ of the particles produced by a cluster are given by
\begin{equation}\langle \mu_n \rangle=\sqrt{\frac{n S_n}{S_1}} \langle \mu_1
  \rangle\ \ \ {\rm and}\ \langle {p_T}^2_n \rangle=\Big ( \frac{n S_1}{S_n}\Big
  )^{1/2} \langle {p_T}^2_1 \rangle \, ,
\label{ec1} \end{equation}
where $\langle \mu_1 \rangle$ and $\langle {p_T}^2_1 \rangle $ are the
corresponding quantities in a single string.  

In the limit of random distribution of strings, eqs. (\ref{ec1}) transform
into:
   \begin{equation} \langle \mu \rangle=N_{S} F(\eta)\langle \mu_1 \rangle \hspace{1cm}
     \langle {p_T}^{2} \rangle=\frac{\langle {p_T}^2_1 \rangle}{F(\eta)}\label{eq:analytic}
     \end{equation} with $F(\eta)=\sqrt\frac{1-\exp{-\eta}}{\eta}$ .

\noindent If we are interested in a
specific kind of particle $i$, we will use $\langle \mu_{1} \rangle_{i}$, $\langle {p_T}^{2}_{1} \rangle_{i}$
$\langle \mu_{n} \rangle_{i}$ and $\langle {p_T}^2_n \rangle_{i}$ for the corresponding quantities.
The transverse momentum distributions can be written as a superposition of the
transverse momentum distributions of each cluster, $g(x,p_{T})$, weighted with
the distribution of the different tension of the clusters, i.e the distribution
of the size of the clusters, $W(x)$ \cite{DiasdeDeus:2003ei} \cite{Cunqueiro:2007fn}.
For  $g(x,p_{T})$ we assume the Schwinger shape, $g(x,p_{T})=\exp(-p_{T}^{2}
x)$ and for the weight function,
$W(x)$, the Gamma distribution $W(x)=\frac{\gamma}{\Gamma(k)}(\gamma x)^{k-1} \exp(-kx)$ with $\gamma=k/\langle x \rangle$ and $k=\langle x \rangle^2/(\langle x^2 \rangle-\langle x \rangle^2)$.
$x$ is proportional to the inverse of the tension of each cluster, precisely
$x=1/\langle {p_T}^{2}_{n} \rangle=\sqrt{\frac{S_{n}}{nS_{1}}}\frac{1}{\langle {p_T}^{2}_{1} \rangle}$. $k$ is
  proportional to the inverse of the width of the distribution on $x$ and
  depends on $\eta$.


The transverse momentum distribution $f(p_{T},y)$ is:
\begin{equation}\frac{dN}{dp_{T}^2 dy}=\int_{0}^{\infty}dx W(x)
g(p_{T},x)= \\
\frac{dN}{dy}\frac{k-1}{k}\frac{1}{\langle {p_T}^2_1 \rangle_{i}}F(\eta)\frac{1}{(1+\frac{F(\eta)p_{T}^{2}}{k\langle {p_T}^2_1 \rangle_{i}})^{k}}
\label{eq:invcs}
\end{equation} 
Eq. (\ref{eq:invcs}) is valid for all densities and types of collisions. It only
depends on the parameters $\langle {p_T}^{2}_{1} \rangle _{i}$.
At low density $\eta$, there is no overlap between strings and therefore 
there are no fluctuations on the cluster size; all the clusters have only 
one string and $k$ goes to infinity. 

\noindent At very high density $\eta$, there is only one cluster formed by 
all the 
produced strings. Again there are no fluctuations, $k$ tends to infinity 
and the transverse momentum distribution recovers the exponential shape. 
In between these two limits, $k$ has a minimum for intermediate densities 
corresponding to the maximum of the cluster size fluctuations. The 
quantitative dependence of $k$ on $\eta$ was obtained from the comparison 
of equation (3) with RHIC $AA$ data at different centralities. The 
peripheral $Au$-$Au$ collisions at RHIC correspond to $\eta$ values 
slightly above the minimum of $k$. Notice, that according to eq. (3) the 
ratio $R_{CP}$ is, at intermediate and high $p_{T}$, proportional to $p_{T}^{2 (k-k')}$, being $k$ and $k'$ the corresponding values for peripheral and central 
collisions respectively. As $k' > k$, $R_{CP}$ is suppressed. 

The equations (\ref{eq:analytic}) and (\ref{eq:invcs}) must be slightly modified in the case
  of baryons to take into account the differences between mesons and baryons
  in the fragmentation of a cluster of strings due to both the
  higher color and the higher possibilities of flavor recombination \cite{Cunqueiro:2007fn}. Due 
  to this, eq. (\ref{eq:analytic}) becomes for (anti)baryons:

\begin{equation}\mu_{\overline{B}}=N_{S}^{1+\alpha}F(\eta_{\overline{B}})\mu_{1\overline{B}}\label{eq:dndy}
\end{equation} where $\alpha \sim$ 0.09 and $\mu_{1\overline{B}} \sim$
0.033$\mu_{1\pi}$. This means that the density $\eta$ must be replaced by $\eta_{\overline{B}}=N_{S}^{\alpha}\eta$.

In figs. 1 and 2 we show the ratio $R_{CP}$ between the inclusive $pp$ going
to $\pi$, $k$ and $\overline{p}$ cross sections for events with a multiplicity
twice larger than the mean multiplicity and the minimum bias cross-sections at
LHC and RHIC energies respectively. The values of the parameters used are 
$\langle {p_T}^2_1 \rangle_{\pi}=$0.06 GeV$^2$/c, $\langle {p_T}^2_1 \rangle_{K}=$0.14
GeV$^2$/c and  $\langle {p_T}^2_1 \rangle_{\overline{p}}=$0.3 GeV$^2$/c. It is
clear why the $p_{T}$ dependence of $R_{CP}$ changes so much when moving from
RHIC to LHC. At RHIC we are still in the low string density regime, with $k$
decreasing with $\eta$. Since $R_{CP} \sim p_{T}^{2 (k-k')}$, and $k 
> k'$, there is no suppression. On the contrary, at LHC for high density 
events we are above the minimum of $k$, $k < k'$ and $R_{CP}$ is 
suppressed.

The values of $\eta$ are computed using for $N_{S}$ the values obtained 
from the quark gluon string model of reference \cite{Amelin:2001sk}, 
whose values are 
similar to the ones obtained using the dual parton model 
\cite{Capella:1992yb} 
or Venus \cite{Werner:1993uh}. Essentially, they are the number of collisions 
times the number 
of strings per collision. This last number is twice the number of pomerons 
in $pp$ collisions at a given energy.


We do not claim to describe all the data including high $p_{T}$. It is well
known that jet quenching is the mechanism responsible for the high $p_{T}$
suppression. This phenomenon is not included in our formula, which was
obtained assuming a single exponential for the decay of a cluster without a
power-like tail. Our formula, must be considered as a way of interpolating and
of joining smoothly the low and intermediate $p_{T}$ region with the  high
$p_{T}$ region.  Indeed the high $p_{T}$ suppression implies by continuity a
suppression of the highest $p_{T}$ values of the intermediate region which are
described by our formula (3). Therefore we have confidence in our results for
low and moderate $p_{T}$, what means that our evaluations are valid for
$p_{T}$ less than $2$-$3$ GeV/c at RHIC energies and $4$-$5$ GeV/c at LHC
energies. Outside this range our evaluations cannot be applied.

The density of partonic matter is at the origin of the observed scaling law of
the elliptic flow, $v_{2}$, normalized to eccentricity ($\epsilon$), that depends only on
the particle density, i.e, the ratio between $dN_{ch}/dy$ and the overlapping
area $S$ \cite{RHIC}, independently of the energy, type and degree of
centrality of the collision. Assuming that this scaling stands also for $pp$ collisions,
we can deduce $v_{2}$ from the observed experimental scaling and from the
values of $\epsilon$,  $dN_{ch}/dy$ and $S$.  Given a degree of centrality, we
compute the impact parameter $b$ using the code of reference
\cite{Amelin:2001sk}. The overlapping area and the eccentricity depend on the
impact parameter and on the nuclear radius $R_{A}$:
$S=2 R_{A}^2 cos^-1(\frac{b}{2R_{A}})-b \sqrt{R_{A}^2-\frac{^2}{4}}$ and
$\epsilon=(\sqrt{2 R_{A}+b}-\sqrt{2 R_{A}-b})/(\sqrt{2
  R_{A}+b})$. $\frac{dN_{ch}}{dy}$ is taken from an extrapolation of $pp$ data
to the LHC in \cite{Abreu:2007kv}.

In any interacting partonic framework, it is natural to assume that the 
scaling law can be extended from $AA$ to $pp$ collisions. In our case, in 
fact, the strings, which are stretched between projectile and target 
partons, interact to form clusters regardless of the initial state 
particles being protons or nuclei. 

The experimental measurement of $v_{2}$
requires high multiplicities but the values of $v_{2}$ corresponding to
central events in $pp$ collisions are quite negligible. In table 1 we show the obtained values of
$v_{2}$ for LHC and RHIC energies for $pp$ minimum bias and also for central
$pp$ collisions at LHC energies assuming the mentioned empirical scaling. We see, that at  LHC for minimum bias the
elliptic flow $v_{2}$ is comparable to the corresponding one for
$Au$-$Au$ collisions at $N_{part}\simeq 250$ \cite{Back:2003hx}. However, the charge
multiplicity is not high enough ($N_{ch} \simeq$  80) \cite{Matinyan:1998ja} what makes the measurement of
$v_{2}$ hard. Needless to say that a higher $v_{2}$ can be obtained for more
$pp$ peripheral collisions, but in this case the charged multiplicity is lower
being the measurement not feasible.\\ A more detailed evaluation of $v_{2}$
done recently \cite{Irais} in the framework of percolation of strings agrees
with the values of table 1.  

On the other hand, one can ask for the possibility of disappearance of back to
back jet-like hadron correlations in $pp$ as in the case of $Au$-$Au$
collisions. Let us assume a hard quark-quark collision produced in the surface
of a proton-proton collision. One outgoing quark will originate the trigger
jet and the other back quark has to go through the 
strings stretched between the projectile and the target partons. The momentum
broadening due to the interaction of the quark with one string has been
computed in \cite{DiasdeDeus:2003ei} to be $$\Delta p_{1}=\frac{4 \pi \alpha_{S}}{3 \sqrt{2}
  r_{0}}\simeq 0.9-0.75  \;\rm{GeV/c}$$
which is taken from the momentum of the quark. The average number of strings
crossed by the quark is $N=2 L \frac{r_{0}}{\pi R_{p}^{2}}N_{S}$ where
$L$ is the distance traveled by the quark $L \simeq R_{p}$. For very central
$pp$ collisions (multiplicity three times larger than the minimum bias
multiplicity) $N_{S} \simeq 40$ and for $r_{0}\simeq 0.2-0.3$ fm we have
$N\simeq 5-8$. As explained in \cite{DiasdeDeus:2003ei}, the total broadening behaves like
$N^{1/2}$, therefore
$$\Delta p_{TOT}=\sqrt{N} \Delta p_{1} \simeq 1.6-2.6 \; \rm{GeV/c}.$$
Therefore the loss is not large, although it would be appreciable for
inclusive particle measurements in the range of $p_{T}\simeq 5-10$ GeV/c.
\\ Recently it has been shown that large long range rapidity correlations are expected to occur in pp at LHC as well as the ridge structure seen at RHIC \cite{Brogueira:2009nj}.
\\Summarizing, in $pp$ collisions at LHC energies, some nuclear like effects
such as high $p_{T}$ suppression, elliptic flow and long range rapidity correlations could occur. Also, some
suppression of the back to back jet like hadron correlations could be
detected. All these effects are consequences of the high density partonic medium.

\section*{Acknowledgments}

We thank N.Armesto and  C. A. Salgado for discussions.
\noindent LC and CP have been supported by MEC of Spain under project
FPA2005-01963, by Xunta de Galicia (Conseller\'ia de Educaci\'on) and by the
Spanish Consolider-Ingenio 2010 Programme CPAN (CSD2007-00042). LC has also been supported by MEC of Spain under a grant of the FPU Program. 

\begin{figure}[]

\centerline{\includegraphics[width=0.8\textwidth,height=0.5\textheight]{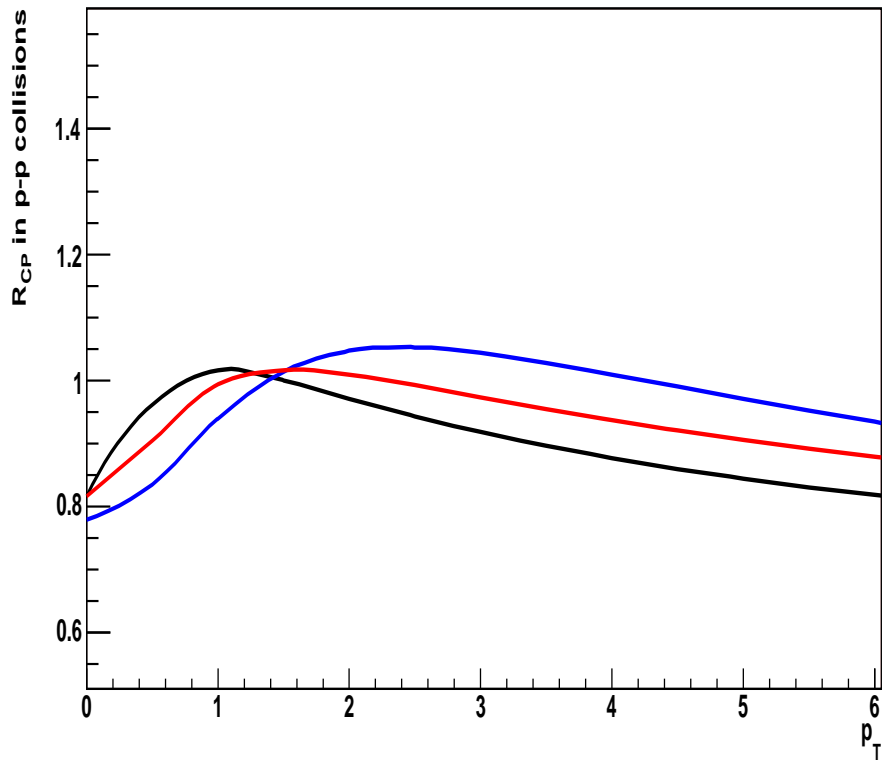}}

\caption{Central to peripheral ratio $R_{CP}$ for pions (black), kaons (red)
  and antiprotons (blue) in $pp$ collisions at LHC energies.}

\label{fig1}

\end{figure}

\begin{figure}[]
\centerline{\includegraphics[width=0.8\textwidth,height=0.5\textheight]{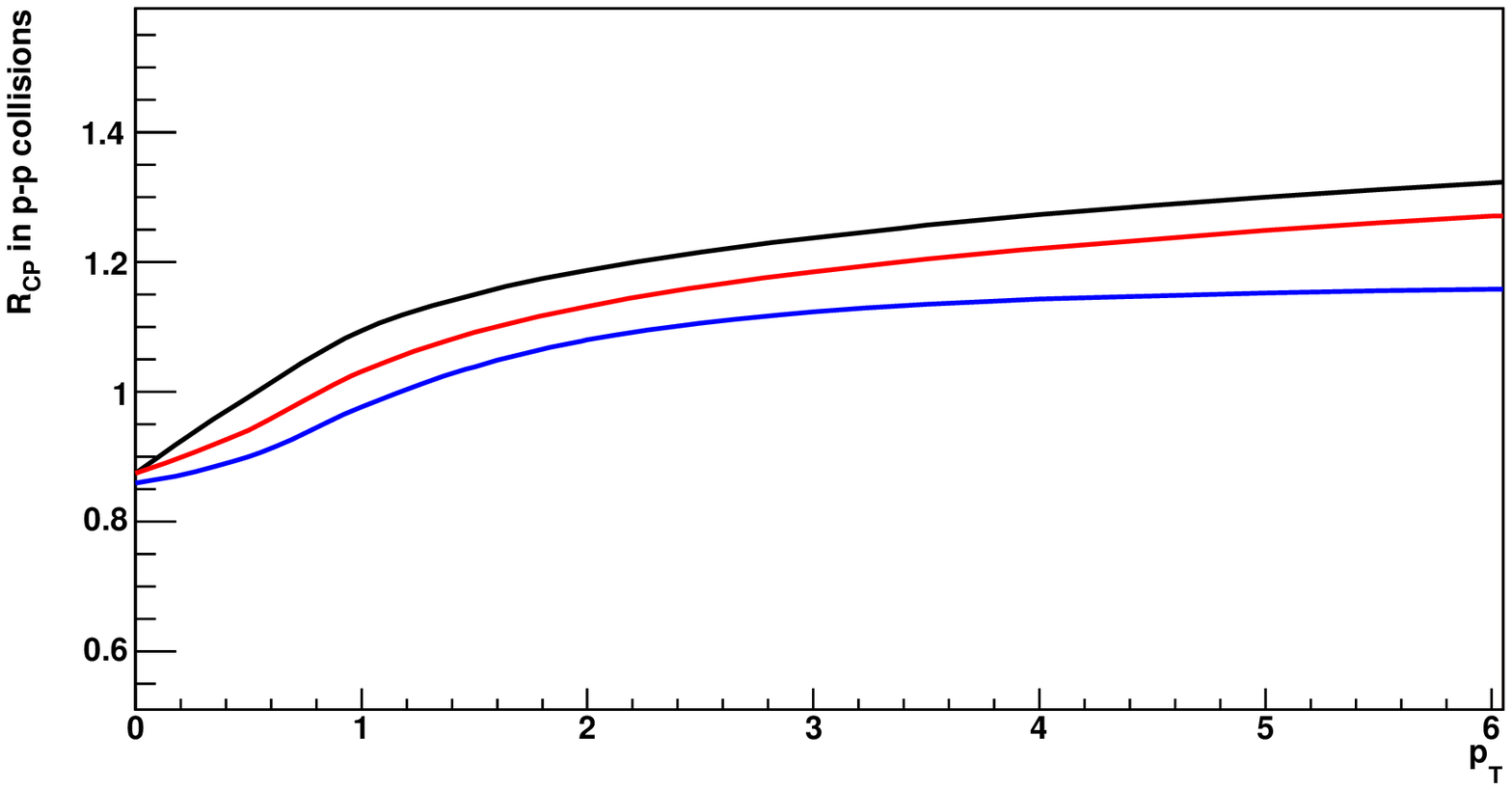}}
\caption{Central to peripheral ratio $R_{CP}$ for pions (black), kaons (red)
  and antiprotons (blue) in $pp$ collisions at RHIC energies.}
\label{fig2}
\end{figure}

\begin{table}
\begin{center}
\begin{tabular}{|c|c|c|c|c|c|c|} \hline
\ \  \ \
&$\sqrt{s}$ \ \

& $b$ (fm)  \ \ & $\epsilon$ \ \ & $S$ (fm$^2$) \ \ & $dN_{ch}/dy$ \ \ & $v_{2}$ \ \       \\ \hline

 $pp$ $0-10\%$  & 14 TeV & 0.36 & 0.17 & 2.42 & 12-15 & 0.008 \\ \hline
 $pp$ min bias  & 14 TeV & 1.18 & 0.49 & 0.93 & 6 & 0.035 \\ \hline
 $pp$ min bias  & 0.2 TeV & 0.8 & 0.35 & 1.58 & 3 & 0.004 \\ \hline
\end{tabular}
\end{center}
\caption{$v_{2}$ in $pp$ collisions for different centralities together with
 the corresponding values for the energy, impact parameter, eccentricity,
 overlapping area and charged central rapidity distribution.}
\end{table}

\end{document}